\documentclass[12pt,preprint]{aastex}
\usepackage{natbib,epsfig}

\begin{document}

\newcommand{\NGC}{\mbox{\protect\small NGC\hspace{.15em}\normalsize}}
\newcommand{\hii}{{\rm H}\,{\sc ii}}
\newcommand{\uchii}{UCH\,{\sc ii}}
\newcommand{\etal}{{\it et. al.}}
\newcommand{\hcoplus}{HCO$^{+}$ ${(1-0)}$}
\newcommand{\co}{$^{12}$C$^{16}$O}
\newcommand{\cyano}{HC$_3$N}
\newcommand{\siomaser}{SiO$(J=1\rightarrow0,\nu=1)$}
\newcommand{\sioone}{SiO$(J=1\rightarrow 0)$}
\newcommand{\siotwo}{SiO$(J=2\rightarrow 1)$}
\newcommand{\siothree}{SiO$(J=3\rightarrow 2)$}
\newcommand{\siofive}{SiO$(J=5\rightarrow 4)$}  
\newcommand{\siofi}{SiO(5-4)}  
\newcommand{\sioeight}{SiO$(J=8\rightarrow 7)$}  
\newcommand{\sot}{SO$_2$}
\newcommand{\nhone}{NH$_3$(1,1)}
\newcommand{\nhtwo}{NH$_3$(2,2)}
\newcommand{\nhthree}{NH$_3$(3,3)}
\newcommand{\ammonia}{NH$_3$}
\newcommand{\methanol}{CH$_3$OH}
\newcommand{\methanolline}{CH$_3$OH $5_{1,4}-4_{2,2}$E}
\newcommand{\methylcyanide}{CH$_3$CN}
\newcommand{\thcoone}{$^{13}$CO ${(J=1\rightarrow0)}$}
\newcommand{\thco}{$^{13}$CO}
\newcommand{\coone}{CO(1-0)}
\newcommand{\cotwo}{CO ${(J=2\rightarrow1)}$}
\newcommand{\ceioone}{C$^{18}$O ${(J=1\rightarrow0)}$}
\newcommand{\ceio}{C$^{18}$O}
\newcommand{\dcn}{DCN(3-2)}
\newcommand{\hcccn}{HCCCN(25-24)}
\newcommand{\hcctcn}{HCC$^{13}$CN(24-23)}
\newcommand{\hctccn}{HC$^{13}$CCN(24-23)}
\newcommand{\htwosline}{H$_2$S $2_{2,0}-2_{1,1}$}
\newcommand{\htwos}{H$_2$S}
\newcommand{\sotwo}{SO$_2 \, 22_{2,20}-22_{1,21}$}
\newcommand{\thfsotwo}{$^{34}$SO$_2 \, 11_{1,11}-10_{0,10}$}
\newcommand{\htwoo}{H$_2$O}
\newcommand{\htw}{H$_2$}
\newcommand{\nh}{NH$_3$}
\newcommand{\msun}{M$_{\sun}$}
\newcommand{\lsun}{L$_{\sun}$}
\newcommand{\cs}{C$^{34}$S ${(J=3\rightarrow2)}$}
\newcommand{\um}{$\mu$m}
\newcommand{\percc}{cm$^{-3}$}
\newcommand{\persqcm}{cm$^{-2}$}
\newcommand{\kms}{km~s$^{-1}$}
\newcommand{\vlsr}{\hbox{${\rm v}_{\rm lsr}$}}
\newcommand{\cstwo}{CS $(J=2\rightarrow1)$ }
\newcommand{\Msun}{M$_\odot$}
\newcommand{\jyperbeam}{\rm{Jy beam$^{-1}$}}
\newcommand{\mjyperbeam}{\rm{mJy beam$^{-1}$}}
\newcommand{\jyperbeamkms}{\rm{Jy Beam$^{-1}$}km~s$^{-1}$}
\newcommand{\arcseconds}{''}
\newcommand{\hthonea}{H$31\alpha$}
\newcommand{\htha}{H$30\alpha$}
\newcommand{\htwsixa}{H$26\alpha$}
\newcommand{\htwonea}{H$21\alpha$}

\title{Mapping the Outflow from G5.89-0.39 in \siofive}

\author{
P. K. Sollins\altaffilmark{1,4}
T. R. Hunter\altaffilmark{1},
J. Battat\altaffilmark{1},
H. Beuther\altaffilmark{1},
P. T. P. Ho\altaffilmark{1,2},
J. Lim\altaffilmark{2},
S. Y. Liu\altaffilmark{2},
N. Ohashi\altaffilmark{3},
T. K. Sridharan\altaffilmark{1},
Y. N. Su\altaffilmark{2},
J.-H. Zhao\altaffilmark{1},
Q. Zhang\altaffilmark{1}
}

\altaffiltext{1}{Harvard-Smithsonian Center for Astrophysics, 60 Garden
Street, Cambridge, MA, 02138}

\altaffiltext{2}{ASIAA, No. 1, Roosevelt Road, Section 4, Taipei 106,
Taiwan, R.O.C.}

\altaffiltext{3}{ASIAA, 645 North A'ohoku Place, Hilo, HI, 96720}

\altaffiltext{4}{psollins@cfa.harvard.edu}

\begin{abstract}

We have mapped the ultracompact \hii\ region, G5.89-0.39, and its
molecular surroundings with the Submillimeter Array at $2\arcseconds.8
 \times 1\arcseconds.8$ angular resolution in 1.3 mm
continuum, \siofive, and eight other molecular lines. We have resolved
for the first time the highly energetic molecular outflow in this
region. At this resolution, the outflow is definitely bipolar and
appears to originate in a 1.3 mm continuum source. The continuum
source peaks in the center of the \hii\ region. The axis of the
outflow lines up with a recently discovered O5V star.

\end{abstract}


\section{Introduction} \label{section:int}

G5.89-0.39 (G5.89) is a shell-like ultracompact (UC) \hii\ region
\citep{woo89b} which was recently found to contain an O5V star via
near-IR imaging \citep{fel03}. \citet{aco98} have used proper motion
measurements of the expansion of G5.89 to determine its distance,
$\sim 2$ kpc, its size, 0.01 pc, and its dynamical age, 600
years. Associated with G5.89 is a molecular outflow \citep{har88} with
a mass of 77 \msun\ and an energy of $5 \times 10^{47}$ ergs
\citep[hereafter, AWC]{aco97}. While many \uchii\ regions have been
observationally associated with molecular outflows
\citep{sne90,she96a}, the question of whether an \uchii\ region can
itself be or contain the source of a bipolar outflow remains in
dispute.  Bipolar outflows are generally understood to be a signpost
of ongoing accretion, while the presence of an \uchii\ region has been
understood by some as a sign that accretion has ceased
\citep{gar99}. Recent observational and theoretical results have shown
that the presence of an \uchii\ region does not necessarily shut off
accretion \citep{ket02a,ket02b}. But since massive stars and their
\uchii\ regions tend to form in very crowded fields, high resolution
imaging is necessary to associate the origin of any particular bipolar
molecular outflow with an \uchii\ region, and not a distinct, nearby
protostar. This determination is possible with and well suited to the
capabilities of the new Submillimeter Array\footnote{The Submillimeter
Array is a joint project between the Smithsonian Astrophysical
Observatory and the Academia Sinica Institute of Astronomy and
Astrophysics, and is funded by the Smithsonian Institution and the
Academia Sinica. For more on the SMA and its specifications see
\citet{ho03}.} (SMA) on Mauna Kea.

We have used the SMA to map the region of G5.89-0.39 in the \siofive\
line, known to trace outflows and this outflow in particular
\citep[AWC]{cod99}. Because of the spectral capabilities of the SMA,
we were simultaneously able to map the region in nine spectral lines
and broadband continuum. The aim of the project was to improve on the
previous $18\arcseconds$ resolution map in \siofi\ of AWC, and also
to detect in dust and molecular lines any other massive protostars in
the vicinity which might be sources of the outflow.

\section{Observations} \label{section:observations}

G5.89-0.39 was observed with the SMA on the night of 2003 13 July with
160 minutes on source. The pointing center of the observations was
$\alpha(2000)=\rm{18^h00^m30^s.32}, \, \delta(2000)=-24^o
04'00''.48$. Five antennas were used in the array, giving baselines
with projected lengths ranging from 10.2 meters to 116 meters
resulting in a $2\arcseconds.8 \times 1\arcseconds.8$ synthesized
beam. The FWHP of the primary beam was $50\arcseconds$. We observed
the quasar NRAO530 for phase calibration every 35 minutes ($\sim
13$~degrees from the source, $S_{217 \rm{GHz}} \sim 2.4$~Jy), and the
data were self-calibrated for phase correction at 1 minute intervals
using the continuum as a model. We observed Mars and Neptune for
bandpass and flux calibration respectively. The average system
temperature was 180~K. We had 1~GHz (1300~\kms) spectral coverage in
each sideband with 0.8125~MHz (1.1~\kms) spectral resolution. The
resulting noise in the 227 GHz (1.3 mm) continuum map was $\sim 20$
\mjyperbeam\ with a bandwidth of 900 MHz. The theoretical noise limit
for this map is 1.7 \mjyperbeam, so the actual noise level is probably
limited by dynamic range. The noise in a single spectral channel map
was $\sim 0.12$ \jyperbeam, which is only a factor of two worse than
the theoretical limit of 0.057 \jyperbeam. We smoothed our data to 3.3
\kms\ resolution for most of the analysis presented below.

\section{Results} \label{section:results}

	\subsection{\siofive\ Map and Other Lines} \label{subsection:lines}

\begin{figure}
\epsscale{1.0}
\plotone{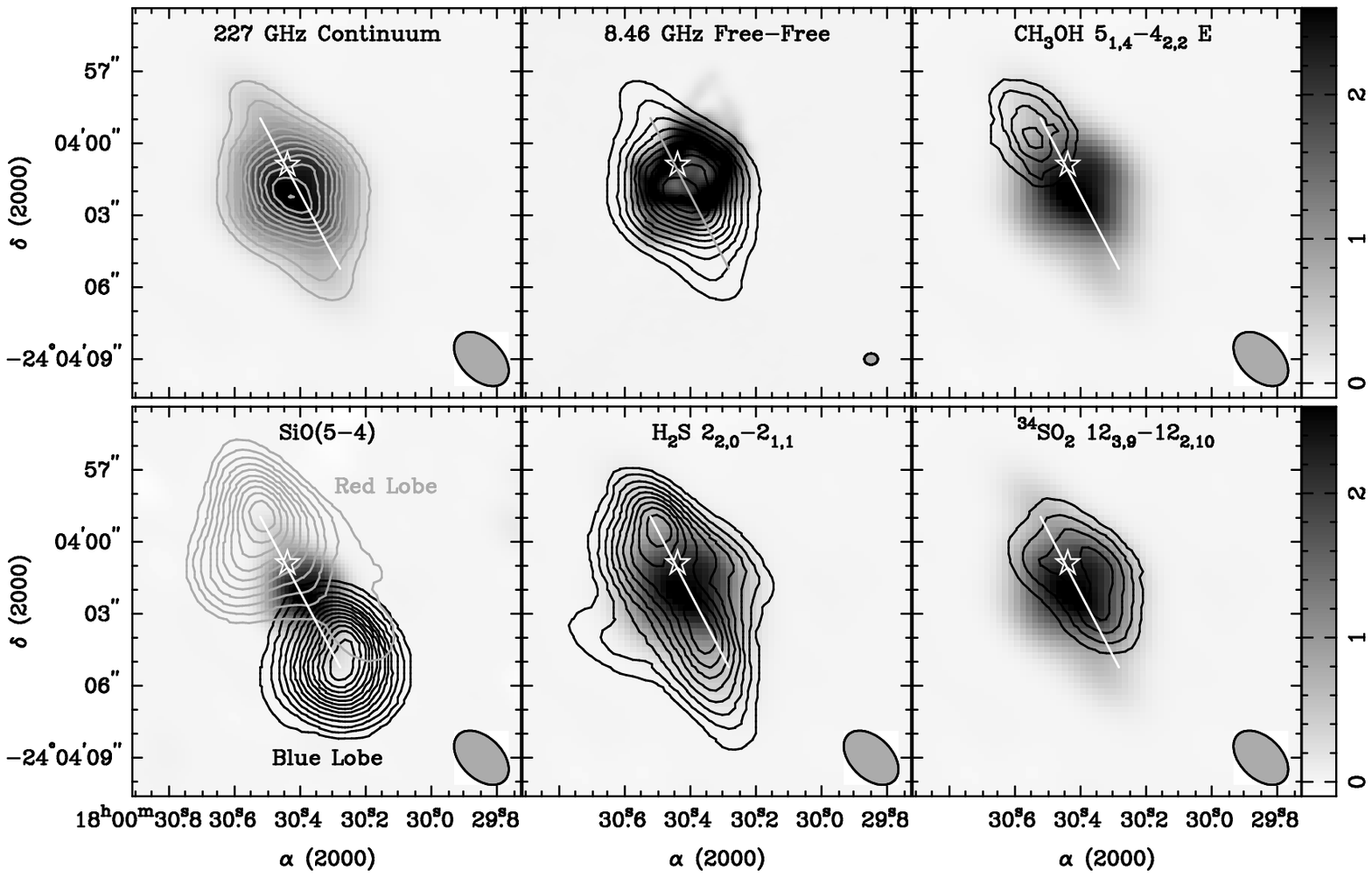}
\caption{227 GHz (1.3 mm) and 8.46 GHz continuum maps are shown in
grayscale, and 227 GHz continuum and integrated line emission maps are
shown in contours. The first panel is 227 GHz continuum in linear
grayscale from -0.1 to 2.6 \jyperbeam\ and contours in multiples of
0.25 \jyperbeam. The second panel shows the same 227 GHz continuum in
contours, and the 8.46 GHz continuum map of \citet{woo89b} in linear
grayscale from -5 \mjyperbeam\ to 100 \mjyperbeam. The remaining
panels all show the 227 GHz continuum in the same grayscale as the
first panel, and contour maps of the velocity integrated brightness in
a variety of molecular lines. The \methanolline\ map is contoured in
multiples of 1 \jyperbeamkms\ and the other three contour maps are
contoured in multiples of 3 \jyperbeamkms. The \siofi\ maps are
integrated from \vlsr\ = -25 \kms\ to 5 \kms\ for the blue lobe and
from \vlsr = 15 \kms\ to 45 \kms for the red lobe. All other lines are
integrated over their full extent in velocity. In all panels the white
line segment connects the positions of the \siofi\ peaks, the white
star represents the position of the O5 star detected by \citet{fel03},
and the beam is in the lower right.\label{figure:linemaps}}
\end{figure}

Our map of the integrated \siofi\ line emission in Figure
\ref{figure:linemaps} clearly resolves and separates the red- and
blue-shifted lobes of the outflow. The extent of the \siofi\ emission
is consistent with the unresolved \siofi\ emission of AWC. We detect
\siofi\ emission which is much stronger and more extended than the
\sioone\ of AWC, probably due to excitation effects in the
outflow. Figure \ref{figure:linemaps} also shows the location of the
O5 star detected by \citet{fel03} and a line connecting the peaks of
the outflow lobes. Both lobes are more than two synthesized beams
across in the direction perpendicular to the outflow axis. At the
peaks of the red- and blue-shifted lobes, the velocity integrated flux
in our \siofi\ spectra are 256~K~\kms and 306~K~\kms,
respectively. Assuming LTE and optically thin emission with a uniform
rotation temperature of 100~K, from AWC's analysis of SiO, CO and
\ammonia, we calculate that the column density of SiO at the peak of
the red- and blue-shifted lobes are $5.4 \times 10^{14}$~\persqcm\ and
$6.4 \times 10^{14}$~\persqcm\ respectively. These are higher than the
column densities determined by AWC, probably because we have much
higher spatial resolution and are able to avoid beam dilution. The
brightest channel in either of the two spectra considered is 16.4~K,
in which channel the optical depth would be less than 0.2, so our
assumptions in this calculation are self-consistent.

\begin{figure}
\epsscale{1.0}
\plotone{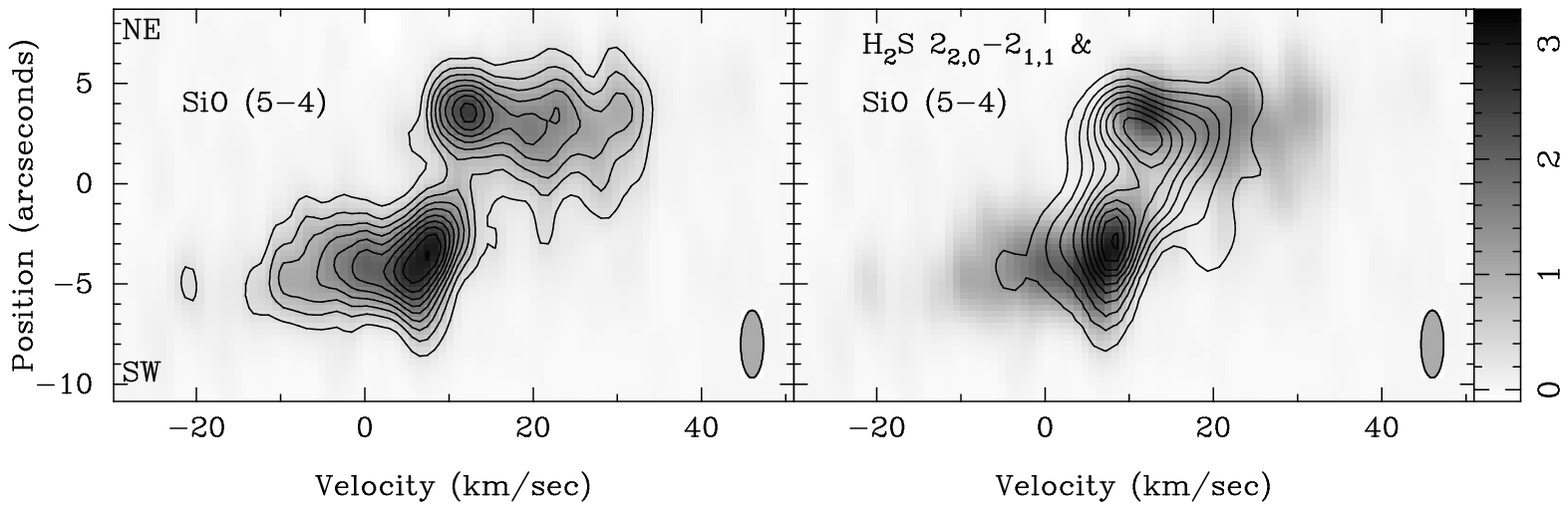}
\caption{A position-velocity diagram of the \siofi\ emission in
grayscale ranging linearly from -0.1 to 3.3 \jyperbeam\ and contoured
at 1 to 10 $\times 0.3$ \jyperbeam\ in the first panel, and in the
second panel, the \siofive\ in the same grayscale and \htwos\ emission
in contours at 1 to 10 $\times 0.3$ \jyperbeam. The data are hanning
smoothed to 3.3 \kms\ spectral resolution, with the resolutions along
both axes shown by the ``beam'' in the lower right. The cut goes
through both of the integrated line emission peaks in the \siofi\ map
in Figure \ref{figure:linemaps} with positive direction in position
being Northeast as shown. \label{figure:posvel}}
\end{figure}

Figure \ref{figure:posvel} shows a position-velocity (P-V) diagram of
the \siofi\ emission. The P-V cut goes through the peaks of the
\siofi\ lobes. There is no obvious dependence of velocity on
position within each lobe, as is seen in some bipolar outflows from
low- to intermediate-mass young stars \citep{lad96,arc01}. The peak
emission is at low relative velocities, $\pm 5$ \kms, and tails off
towards higher relative velocities. The \siofi\ emission exceeding
$3\sigma$ ranges from \vlsr = -20 \kms\ to +40 \kms. Red- and
blue-shifted \coone\ and \hcoplus\ emission have been detected out to
$\sim 30\arcseconds$ from the \uchii\ region \citep{wat02}, but the
\siofi\ emission extends only $\sim 5\arcseconds$ in either
direction along the outflow giving a dynamical time-scale of 1600
years (separation divided by velocity). This number, however, is
uncertain, since the line-of-sight velocity is a lower limit on the
three dimensional velocity and the projected separation is a lower
limit on the three dimensional separation.

In total we detected 9 spectral lines. \siofi, \htwosline, \hcccn,
\sotwo, and \thfsotwo, all show the outflow to varying degrees.
\hctccn, \hcctcn, and \methanolline, show only marginally resolved
emission all at the same location in the Northeast lobe of the outflow
and all at the ambient velocity of \vlsr=10 \kms. \dcn\ was also
detected but was difficult to map, probably due to missing flux.

	\subsection{227 GHz Continuum} \label{subsection:continuum}

The first panel of Figure \ref{figure:linemaps} shows our 227 GHz
continuum emission map. 900 MHz of line-free channels from the upper
sideband were averaged to obtain the continuum visibilities. The map
shows the full extent of the detected 1.3 mm continuum emission. We
detect no other 1.3 mm continuum sources anywhere in our
$50\arcseconds$ field of view down to a $3\sigma$ detection limit of
60 mJy. The second panel in Figure \ref{figure:linemaps} shows the
8.46 GHz continuum map of \citet{woo89b} in grayscale overlaid with
our 1.3 mm map in contours, and the stellar source of
\citet{fel03}. The peak in 1.3 mm emission is about $1\arcseconds.5$
removed from the star. The strongest 1.3 mm emission is extended
nearly perpendicular to the outflow axis. The faint extensions of the
1.3 mm emission follow the axis of the outflow and are probably
emission from dust in the outflowing molecular gas, as has been
observed in L1157 \citep{gue03}.

\begin{figure}
\epsscale{1.0}
\plotone{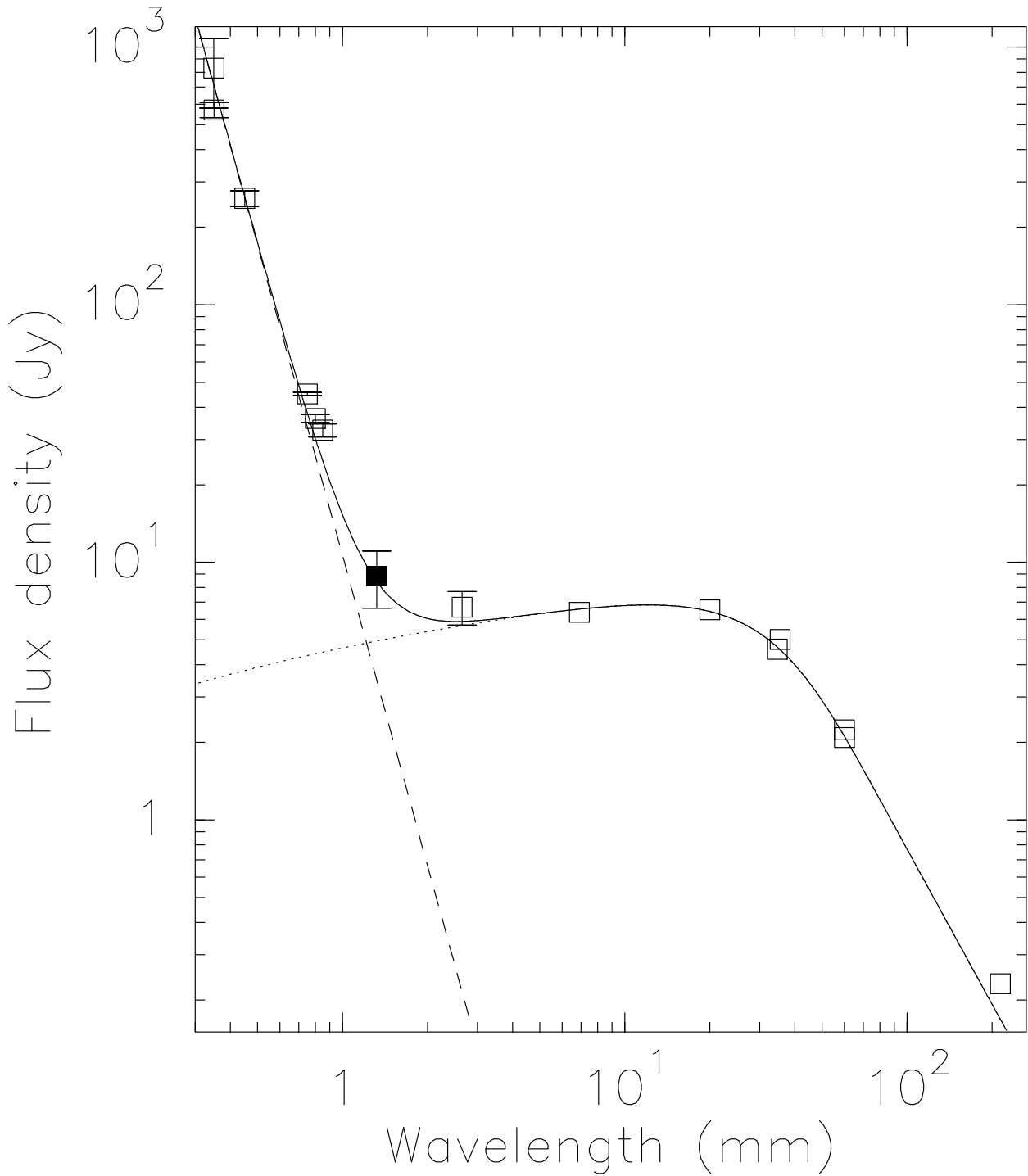}
\caption{The spectral energy distribution of G5.89-0.39. Points shown
are from the VLA and ATCA in the free-free part of the spectrum, and
from the SMA, OVRO, JCMT and CSO in the dust dominated part of the
spectrum. The fit to the free-free spectrum is shown as a dotted
curve, the fit to the dust emission is dashed, and the sum of the two
is solid. Error bars are shown were they are known. The fits include
data from \citet{ake96,san94,aco98,bec94,woo89b,hun97}. The datum from
this work is a solid square. Other data are open
squares. \label{figure:sed}}
\end{figure}

The total flux detected in the 1.3 mm map is 8.8 Jy, integrated over a
central square of $16\arcseconds \times 16\arcseconds$. Continuum data
from 21 cm to 350 \um\ are plotted in Figure \ref{figure:sed} showing
the SED of this region. Using the distance of 2.0 kpc determined by
\citet{aco98}, we did a least squares fit of a standard free-free
spectrum \citep{gor88} to the data below 50 GHz. We get an electron
temperature of 8000 K, angular diameter of $3\arcseconds.1$
corresponding to a physical size of 6000 AU (0.03 pc), an electron
density of $1.4 \times 10^5$ \percc, and an emission measure of $6
\times 10^8$ pc cm$^{-6}$. For the dust-dominated, short-wavelength
regime, we simply extended the existing ``cold-component'' fit from
\citet{hun00}. The sum of the two spectra is the solid line in Figure
\ref{figure:sed} and fits our data point well. Based on these fits, we
predict 5.6 Jy from free-free emission and 3.2 Jy from dust
emission. Following \citet{hun00}, assuming a dust emissivity of
$Q(\nu=227 \rm{GHz}) = 2.0 \times 10^{-5}$, a dust temperature of 100
K as above, and a total mass to dust mass ratio of $M_{total}/M_{dust}
= 100$ \citep{tes98}, we calculate a total mass of 83 \msun\
associated with the dust emission. If we define the 2 \mjyperbeam\
contour of the free-free map as a boundary, the 1.3 mm map contains
8.2 Jy of flux inside the boundary and 0.7 Jy outside. Since the flux
outside is largely from the extensions which follow the outflow, we
estimate the mass of gas associated with the dust in the outflow to be
at least 18 \msun. This number is an approximate lower limit since
much of outflow overlaps with the \uchii\ region, but since the
effects of the outflow on abundance are unknown it is not a hard lower
limit. The remaining 65 \msun\ is projected against the \uchii\
region.

\section{Discussion} \label{section:discussion}

We believe the origin of the outflow is the 1.3 mm continuum source
for two reasons. First, the slightly extended 1.3 mm continuum peak
lies on the projected outflow axis, halfway between the peaks of the
two outflow lobes. Second, the 1.3 mm continuum source seems to show a
physical connection to both outflow lobes in the form of the
extensions of the continuum along the outflow axis. Apparently the
dust is being swept up in both lobes of the outflow (as seen in L1157
by \citet{gue03}), and thus it seems that the outflow is emanating
from the 1.3 mm continuum source.

The question of whether the outflow actually originates {\it inside}
the \uchii\ region is somewhat thornier. One possibility is that the
1.3 mm continuum source lines up with the \uchii\ region by chance,
and is a distinct high-mass protostar with its own dusty envelope, not
physically connected to the \uchii\ region, but part of the same young
cluster. Alternatively, the dust and free-free continuua could be from
a single source, one dusty, hot molecular core surrounding an embedded
\uchii\ region.  In this case, it seems the outflow would have to come
from inside the \uchii\ region in order for both sides of the outflow
to be seen in the dust continuum. We cannot distinguish between these
two cases, and neither seems unlikely. However, we can say that the
known O5 star is a bad candidate to be the source of the outflow since
it is not equidistant from the outflow lobes.

We cannot eliminate any model chronologies by comparing
timescales. The outflow has a dynamical age of 1600 years, compared to
just 600 for the \uchii\ region. This might lead one to believe that
the outflow is truly older than the \uchii\ region. However, any
``quenching'' of the \uchii\ region by infalling material would tend
to make the dynamical age of the \hii\ region an underestimate of its
true age, perhaps by a great deal \citep{ket02b}. The gas in the
outflow was probably accelerated before the \uchii\ region began to
expand, but not necessarily before the \uchii\ region was created.

The central unknown about the outflow is its inclination. The fact
that the two lobes hardly overlap at all on the plane of the sky and
the fact that the \siofi\ emission is extended perpendicular to the
outflow axis might lead one to believe that the outflow has a wide
opening angle, and lies primarily in the plane of the sky. But these
two facts could just as easily be interpreted as a highly collimated
outflow, like those seen around low-mass proto-stars \citep{bac91},
whose axis is largely along the line of sight. Including only the
line-of-sight velocity, the outflow is already quite energetic, so
adding a large component in the plane of the sky would make this a
remarkable outflow, indeed. Some studies of regions of massive star
formation have found poor collimation of outflows
\citep{ric00,rid01}. But \citet{beu02a} found that 15 outflows with
apparently low collimation were consistent with the very high
collimations seen in low mass cases, with the discrepancy due to low
resolution. Higher resolution interferometer studies have confirmed
the presence of highly collimated outflows in some regions of
high-mass star formation \citep{beu02b}.

The velocity structure of the \siofi\ is intriguing and may be a clue
as to the outflow orientation. In both lobes, the spectra peak at low
relative velocity and tail off towards much higher relative
velocity. In the case of outflows from low-mass protostars, the SiO
emission is concentrated immediately behind the shock in the highest
velocity molecular gas \citep{van98}. So perhaps the fact that the low
relative velocity peaks in the \siofi\ emission, only $\pm 5$ \kms, is
an indication that the outflow is mainly in the plane of the sky. This
outflow, however, is far more energetic than an outflow from a
low-mass protostar. There could be highly excited SiO emission which
follows behind the shock more closely than the \siofi\ emission. We
plan to observe \sioeight, and lines of several CO isotopes with the
SMA allowing calculations of excitation and abundance in the outflow
to see if higher excitation gas occurs at higher line of sight
velocities, as we might expect for an outflow largely along the line
of sight. These observations will also have higher angular resolution
which should clarify the outflow opening angle.

On an equally hypothetical note, consider the fact that while some
dust emission appears to come from the outflow, the free-free emission
shows no preference for the direction of the outflow, and is more
extended perpendicular to the outflow. Recent results have shown that
massive protostars ($\sim 10$\msun) can have disk-like structures akin
to the disks seen in low-mass protostars \citep{zha98b,zha02}. If such
a disk were to photoevaporate, as was modeled by \citet{hol94}, that
would naturally produce free-free emission which is extended
perpendicular to the outflow, and possibly ring-like. However, if the
outflow does not originate in the \uchii\ region, the more traditional
limb-brightened-shell interpretation of the free-free emission may
make more sense. Future work to investigate the geometry and the
velocity structure of the ionized gas could be carried out with the
SMA using low quantum number radio recombination lines (RRLs) of
hydrogen, including \htwonea, \htwsixa, \htha, and \hthonea. These
lines suffer less from pressure broadening than higher quantum number
RRLs, and could be used to probe the velocity structure of the densest
ionized gas in the \uchii\ region.

\bibliographystyle{apj}
\bibliography{bib_entries}

\end{document}